%% file: 0_main.tex
\documentclass[runningheads]{llncs}
\usepackage[sectionbib]{natbib}
\setcitestyle{authoryear}

\usepackage{amsthm}
\usepackage[english]{babel}
\usepackage[misc,geometry]{ifsym}
\usepackage{tikz}
\usepackage{graphicx}
\graphicspath{{./img/}}
\usepackage[T1]{fontenc}
\usepackage{nth}
\usepackage{csquotes}
\usepackage{xfrac}
\usepackage{amssymb}
\usepackage{algorithm2e}

\usepackage{subcaption}
\usepackage{silence}
\usepackage{mathtools}
\usepackage{dsfont}
\usepackage{todonotes}
\usepackage{hyperref}
\newcommand{\1}{\mathds{1}}

\renewcommand{\iff}{\Leftrightarrow}
\newcommand\R{\mathbb{R}}

\newcommand{\argmax}{\operatorname{arg\,max}}

\title{The Optimality of Upgrade Pricing}
\date{\today}
\author{Dirk Bergemann\inst{1} \and Alessandro Bonatti\inst{2} \and 
Andreas Haupt\Letter\inst{3} \and Alex Smolin\inst{4}\orcidID{0000-0003-4740-2376}}
\authorrunning{D. Bergemann, A. Bonatti, A. Haupt, A. Smolin}
\institute{Yale University, Department of Economics, New Haven, USA\\ \email{dirk.bergemann@yale.edu}\and Massachusetts Institute of Technology, Sloan School of Management, Cambridge, USA\\ \email{bonatti@mit.edu} \and Massachusetts Institute of Technology, Institute for Data, Systems, and Society, Cambridge, USA\\ \email{haupt@mit.edu} \and Toulouse School of Economics, University of Toulouse Capitole, Toulouse, France\\ \email{alexey.v.smolin@gmail.com}}
\allowdisplaybreaks
\begin{document}
\maketitle
\begin{abstract}
We consider a multiproduct monopoly pricing model. We provide sufficient conditions under which the optimal mechanism can be implemented via upgrade pricing---a menu of product bundles that are nested in the strong set order. Our approach exploits duality methods to identify conditions on the distribution of consumer types under which  (a) each product is purchased by the same set of buyers as under separate monopoly pricing (though the transfers can be different), and (b) these sets are nested.
 
We exhibit two distinct sets of sufficient conditions. The first set of conditions weakens the monotonicity requirement of types and virtual values but  maintains a \emph{regularity} assumption, i.e., that the product-by-product revenue curves are single-peaked.   The second set of conditions establishes the optimality of upgrade pricing for type spaces with \emph{monotone marginal rates of substitution (MRS)}---the relative preference ratios for any two products are monotone across types. The monotone MRS condition allows us to relax the earlier regularity assumption. 

Under both sets of conditions, we fully characterize the product bundles and prices that form the optimal upgrade pricing menu. Finally, we show that, if the consumer's types are monotone, the seller can equivalently post a vector of single-item prices: upgrade pricing and separate pricing are equivalent.

\end{abstract}
\clearpage
\input{1_intro}
\input{2_model}
\input{3_regular}
\input{4_ironing}
\input{5_separate}
\input{7_conclusion}
\subsubsection*{Acknowledgements}
We thank Mark Armstrong and the seminar audience at MIT for helpful comments. Bergemann and Bonatti acknowledge financial support through NSF SES 1948336. Smolin acknowledges funding from the French
National Research Agency (ANR) under the Investments for the Future (Investissements d’Avenir) program (grant
ANR-17-EURE-0010).
\bibliographystyle{abbrvnat}
\bibliography{references}
\end{document}

%% file: 1_intro.tex
\section{Introduction}
\subsection{Motivation and Results}
Pricing multiple goods with market power is a canonical problem in the
theory of mechanism design. It is also a challenge of growing importance and complexity for online retailers and service providers, such as Amazon and Netflix. Both in theory and in practice, designing the optimal \emph{mixed bundling} mechanism, (i.e., pricing every subset of products) becomes exceedingly complex in the presence of a large number of goods.

A natural question is then whether simpler pricing schemes are optimal under suitable demand conditions. A simple, commonly used  mechanism consists of \emph{upgrade pricing}, whereby the available options are ranked by set inclusion, i.e., some goods are only available as add-ons, \cite{Ellison2005}. For example, many online streaming services use a tiered subscription model, whereby users can pay to upgrade to a \enquote{premium package}---a subscription with a larger selection of the provider's content relative to the \enquote{basic package}, \cite{phil17}.

In this paper, we obtain sufficient conditions under which upgrade pricing maximizes the seller's revenue. Our approach consists of first identifying conditions under which the consumer's types can be ordered in terms of their absolute or relative willingness to pay for the seller's goods, and then ranking the goods themselves by the profitability of selling them to larger sets of consumer types. Our sufficient conditions not only establish the optimality of \emph{some} upgrade pricing menu: they  also show that the optimal bundles are deterministic, and they reveal the order in which they are ranked in the menu. That is, we identify all the nested bundles that appear in the seller's menu, and the profit-maximizing price for each one.

Our results consist of two distinct sets of conditions. The first set of conditions  (Theorem \ref{thm:reg}) illustrates the essence  upgrade pricing optimality in what we label as \textquotedblleft regular\textquotedblright\ settings. While these conditions are reminiscent of regularity in one dimension,  they are in fact  weaker than the monotonicity of the buyer's multidimensional types and of the  (item by item) Myersonian virtual values. What we require is for the consumer's types to be ranked in such a way that the virtual values for each item are negative over an initial and positive over a final segment. Furthermore, we require any consumer with a positive virtual value for an item to also have a larger value for that item, relative to any type with a negative virtual value. At the optimal prices, the lowest type buying each good is indifferent between buying it and not buying it. Finally, the sets of types buying each item are nested under the \emph{weak} monotonicity property, which implies the optimal  allocation can be implemented via upgrade pricing.

The second set of conditions (Theorem \ref{thm:mrs}) describes our best attempt at  extending our approach to non-regular distribution of types. In order to further weaken the regularity requirement, we restrict attention to type spaces for which the relative preference ratios for any two goods are monotone across types. An example of ordered relative preferences is if higher types have a stronger preference for good 2 over good 1. We refer to such a condition as \enquote{monotone marginal rates of substitution} (monotone MRS).

The intuition for our two results can be grasped by considering the demand functions for each good separately. Under monotonicity and monotone MRS, the optimal monopoly prices for each of the goods are ranked. In the special case where the Myersonian virtual values for our ordered types
\[
\phi _{i}^{k}=\theta _{i}^{k}-\frac{1-F_{i}}{f_{i}}\left( \theta_{i+1}^{k}-\theta _{i}^{k}\right)
\]
are also monotone for each item $k$, the first set of conditions applies. 

When virtual values are not monotone, however, they can cross zero more than once. In that case, the result still holds, but the proof requires the right ironing procedure. Our ironing procedure relaxes the standard approach of \cite{myer81} and the literature up to \cite{Haghpanah2020}. Specifically, we do not iron with the goal of monotone virtual values, which corresponds to a concave revenue curve. Rather we iron towards single-crossing virtual values which leads to a \emph{quasi}concave revenue curve. We then use the structure implied by monotone MRS to derive a dual certificate of optimality.

Under either set of conditions, each good is purchased by the same set of buyers that would buy it if that were the seller's \emph{only} product. We further show (Theorem \ref{thm:sep}) that, if the consumer's types are (not weakly) monotone, the seller can equivalently post the vector of single-item monopoly prices---i.e., bundling is redundant. For example, in the case of two goods sold separately, monotone type spaces mean that no consumer type will buy good $2$ without also buying good $1$. More generally, the seller benefits from restricting the set of bundles the consumer can purchase through a proper menu of options with the upgrade property. However, examples also show that implementability through separate pricing is neither necessary nor sufficient for the optimality of upgrade pricing.

\subsection{Related Literature}
First and foremost, our paper contributes to the economics literature on product bundling. The profitability of mixed bundling relative to separate pricing was first examined by \cite{Adams1976}, and further generalized by \cite{mmw89}. More recently, a number of contributions have studied the optimal selling mechanisms in the case of two or three goods, and derived conditions for the optimality of pure bundling (see, for example, \cite{mavi06} and \cite{pav11a}). \cite{Daskalakis2017} use duality methods to characterize the solution of the multiproduct monopolist's problem, and show how the optimal mechanism may involve a continuum of lotteries over items. \cite{bik20} derive conditions under which the optimal mechanism is deterministic when the buyer's utility is not necessarily additive. Finally, \cite{ghili21} establishes conditions for the optimality of pure bundling when buyers' values are interdependent. Relative to all these papers, we focus on a specific class of simple mechanisms, which includes pure bundling as a special case.

\cite{hani17} and \cite{Babaioff2014} also study the properties of simpler schemes. The former derives a lower bound on the revenue obtained from separate item pricing. The  latter obtains an upper bound on the revenue of the optimal mechanism, relative to the better of pure bundling and separate pricing.

In the context of nonlinear pricing, \cite{wils93} suggested a ``demand profile'' approach that determines the price of each incremental unit by treating it as separate market. This approach is particularly attractive in settings where there is a natural ordering over the items. This in particular is the case when there is a homogeneous good that is offered in various quantities, such as in energy markets for electricity or water. This approach naturally generates a sequence of upgrade prices. The demand profile approach, and in particular the incremental pricing rule implied by it, does not always yield an optimal mechanism as consumers may wish to obtain earlier units in order to obtain the later units. Thus, a contribution of the current paper is to determine when upgrade pricing is exactly optimal and then to find the upgrade prices as solutions to the global revenue maximization problem rather than the incremental item problem.  Other papers make assumptions that make sure that a demand profile-type approach yields an optimal mechanism. In \cite{jomy03}, buyers have unit demand and sellers offer different varieties of a single good. The approach in their paper is to assume a quality ranking on the varieties and to solve for the upgrade prices---the additional payments required to buy a better variety. The survey of the nonlinear pricing literature by \cite{armstrong2016nonlinear} covers related approaches that optimize upgrades separately.

Our formulation of the dual problem follows \cite{Cai2016a}, who present a general duality approach to Bayesian mechanism design. \cite{Cai2016a} formulate virtual valuations in terms of dual variables, state the weak and the strong duality results, and use them to establish lower bounds for relative performance of simple mechanisms. An important contribution by \cite{Haghpanah2020} exploits the duality machinery to provide sufficient conditions for the exact optimality of a specific, simple mechanism---pure bundling---consisting of offering a maximal bundle at a posted price. Under their sufficient conditions, the dual variables can be recovered from a single-dimensional problem in which the seller is restricted to bundle all items together.

We follow the approach of  \cite{Haghpanah2020} by leveraging the duality approach to provide sufficient conditions for the optimality of a particular class of mechanisms.  \cite{Haghpanah2020} gave a characterization of the optimality of the grand bundle, we  provide a characterization for upgrade pricing. As upgrade pricing allows multiple bundles to be present in the menu, we cannot assign the dual variables by solving a one-dimensional problem. Instead, we develop a novel ironing algorithm that generates these variables by ironing different item's revenue curves for different types. Under our sufficient conditions, the so-constructed virtual surplus is maximized by an element-wise monotone allocation that can be implemented by upgrade pricing; by complementary slackness, this certifies the optimality of upgrade pricing. Because pure bundling is one instance of upgrade pricing, our conditions differ from those of \cite{Haghpanah2020}.

Our ironing differs from existing ironing approaches using duality and tackles a more general problem. In comparison to \cite{Haghpanah2020}, we prove optimality for mechanisms with menu size surpassing two. \cite{fiat2016fedex} studies a two-parameter model, and uses an ironing approach that leads from the revenue curves to their concave closure. \cite{devanur2020optimal} generalizes \cite{fiat2016fedex} to more general orders on the second parameter. Our approach tackles optimality for an arbitrary finite number of items and varies the ironing procedure. On a technical level, our ironing procedure yields  quasi-concave ironed revenue curves, whereas the ironed revenue curves in \cite{Haghpanah2020,fiat2016fedex,devanur2020optimal} are concave.

Our results also feed into a literature specifying optimal finite mechanisms for multi-dimensional types. \cite[section 7]{Daskalakis2017} for example characterizes the optimal mechanisms for the two-good monopolist problem if the optimal mechanism has a particular structure. While \cite{Daskalakis2017} requires that the region of the type space that is not allocated any item is not adjacent to all regions getting specific constant allocations, upgrade pricing mechanisms consistently break this requirement.

\subsection{Structure of the Paper} The model is introduced in \autoref{sec:model}. The first set of sufficient condition is presented in \autoref{sec:weakreg}. In \autoref{sec:ironing}, we present our results for monotone MRS type spaces. In \autoref{sec:sep}, we discuss the relationship between separate pricing and upgrade pricing. 
We conclude in \autoref{sec:conclusion}.

%% file: 2_model.tex
\section{Model}\label{sec:model}
We consider a standard multiple-good monopoly setting. There is a single seller of $d\geq 1$ goods and a single buyer. The seller's marginal costs of production are normalized to zero. The buyer's utility function is additive across goods. We refer to the vector of marginal utilities $\theta_i\in\R^d$ as the buyer's type. Therefore, the utility of buyer type $\theta_i$ from the consumption vector $q\in[0,1]^d$ is given by
\[
U(\theta_i,q)=\sum_{k=1}^d \theta_i^k q^k.
\]
We also adopt the  shorthand notation $\langle \theta_i, q \rangle \coloneqq \sum_{k=1}^d \theta_i^k q^k$. As a convention, we denote types by subscripts and items by superscripts. The buyer's utility is quasi-linear in transfers and her outside option is also normalized to zero.

The buyer knows her type. From the seller's perspective, the buyer's type is distributed over a finite set $\Theta\subseteq\R^d_+$, with $\lvert\Theta\rvert=n$, according to the distribution $f\in \Delta (\Theta)$. For any positive integer $n$, we adopt the convention that $[n] \coloneqq \{1, 2, \dots, n\}$, and we index types by $i \in  [n]$. We let $f_i \coloneqq f(\theta_i)$ and  denote the cumulative distribution sequence by $F_i = \sum_{j=1}^i f_j$, $i \in [n]$.

The seller aims to maximize revenue. By the revelation principle, we can focus on direct mechanisms $(q,t)=(q_i,t_i)_{i\in \{0\} \cup [n]}$. These mechanisms can be  interpreted as menus with $n+1$ items so that item $i$ delivers consumption vector $q_i$ at price $t_i$ and item $(q_0, t_0) \coloneqq (0,0)$ captures the buyer's  outside option.

We call a menu \textit{upgrade pricing} if $\{q_0, q_1, \dots, q_n\}$ can be ordered in the component-wise partial order on $\R^d$ given by $q \le q' \iff \forall k \in [d]\colon q^k \le (q')^k$. Our main goal is to provide conditions under which upgrade pricing maximizes the seller's revenue among all direct mechanisms.

%% file: 3_regular.tex
\section{Optimal Mechanisms for Regular Distributions}\label{sec:weakreg}

We will make prominent use of the (partial) Lagrangian duality-based certificate of optimality used by \cite{Cai2016a}. We state the underlying duality result to fix notation.
\subsection{Duality}
In what follows, we will associate with $\lambda_{ji}$ the Lagrange multiplier of the incentive compatibility constraint of type $\theta_j$ deviating to type $\theta_i$, $j \in [n], i \in \{0\} \cup [n]$:
\[
	\langle q_j ,  \theta_j \rangle - t_j \ge \langle q_i , \theta_j \rangle - t_i.
\]
We note that the incentive constraints corresponding to $\lambda_{j0}$, $j \in [n]$ are type $j$'s individual rationality constraints. As a main tool in our analysis, we define the multi-dimensional \emph{virtual values} associated with Lagrange multipliers $\lambda\in\R^n\times\R^{n+1}$ as
\begin{equation}
\phi_i^\lambda \coloneqq \theta_i - \frac{1}{f_i} \sum_{j=1}^n \lambda_{ji} (\theta_j - \theta_i).\label{eq:virtvalue}
\end{equation}
\begin{lemma}\label{lem:lagrange}
	A mechanism $(q_i,t_i)_{i \in \{0\} \cup [n]}$ maximizes revenue if and only if there exist multipliers $\lambda_{ji}$, $j \in [n]$, $i \in \{0\} \cup [n]$ such that
	\begin{enumerate}
	\item $\lambda_{ji} \ge 0$ (Non-Negativity)
	\item\label{enum:max} $(q_i)_{i \in [n]}$ optimizes $\max_{(q_i)_{i \in [n]} \in [0,1]^n}\sum_{i=1}^n f_i \langle q_i \cdot \phi_i^\lambda \rangle$ (Virtual Welfare Maximization)
	\item\label{enum:flow} $f_i = \sum_{j=0}^n \lambda_{ij} -  \sum_{j=1}^n\lambda_{ji}$ for all $i \in [n]$ (Feasibility of Flow)
	\item\label{enum:slack} $\lambda_{ji} (\langle q_j , \theta_j\rangle - t_j - \langle q_i , \theta_j\rangle - t_i) = 0$ for all $j\in [n],i \in \{0\} \cup [n]$ (Complementary Slackness)
	\item \label{enum:feas}There are transfers $t$ such that $(q,t)$ is incentive compatible and individually rational (Implementability)
\end{enumerate}
\end{lemma}
We call the dual variables $\lambda_{ji}$, $j \in [n]$, $i \in [n] \cup \{0\}$ \emph{flows} from type $j$ to type $i$ whenever they are non-negative and satisfy \autoref{lem:lagrange} \autoref{enum:flow}. This name is inspired by flow conservation constraints from the  maximum flow and minimum cost flow problem in discrete mathematics \citep{korte2011combinatorial}. 

\begin{proof}[Proof of \autoref{lem:lagrange}]
	Slater's condition for affine inequality constraints \cite[p. 227]{Luethi2005ab} allows us to write revenue maximization subject to the incentive compatibility and individual rationality constraints as an unconstrained optimization problem for $(q_i, t_i)_{i\in [n]}$ subject to complementary slackness and non-negativity of dual variables. The Lagrangian reads:
\begin{align*}
	\mathcal L &= \sum_{i=1}^n f_it_i + \sum_{j=1}^n\sum_{i=0}^n \lambda_{ji} (\langle q_j, \theta_j  \rangle - t_j -\langle  q_i, \theta_j\rangle - t_i) \\
	&= \sum_{i=1}^n t_i \left(f_i - \sum_{j=0}^n \lambda_{ij}   + \sum_{j=1}^n\lambda_{ji}\right) 
	+ \sum_{j=1}^n\sum_{i=0}^n \lambda_{ji} \langle q_j, \theta_j  \rangle - \sum_{j=1}^n\sum_{i=0}^n \lambda_{ji} \langle q_i, \theta_j  \rangle\\
	&=  \sum_{j=1}^n\sum_{i=0}^n \lambda_{ji} \langle q_j, \theta_j  \rangle - \sum_{j=1}^n\sum_{i=0}^n \lambda_{ji} \langle q_i, \theta_j  \rangle\\
	&= \sum_{j=1}^n\left( \left(\sum_{i=1}^n \lambda_{ij}   - \sum_{i=0}^n\lambda_{ji} \right) \langle q_j, \theta_j\rangle - \sum_{i=0}^n \lambda_{ji} (\langle q_i , \theta_j\rangle - \langle q_j , \theta_j\rangle )\right)\\
	&= \sum_{j=1}^n \left( f_j \langle q_j, \theta_j\rangle - \sum_{i=0}^n \lambda_{ji} (\langle q_i , \theta_j\rangle - \langle q_j , \theta_j\rangle )\right)= \sum_{j=1}^n f_j \langle q_j , \phi_j\rangle.
\end{align*}
Clearly, it is necessary for an optimal mechanism to be implementable. To conclude the proof, we need to show that virtual welfare maximization and feasibility of flow are equivalent to maximizing the Lagrangian. Assume virtual welfare maximization and feasibility of flow. Then the above equalities show that the Lagrangian is maximized and certify the optimality of the mechanism $(q_i, t_i)_{i \in [n]}$. Conversely, assume that the Lagrangian is maximized by $(q_i, t_i)_{i \in [n]}$. If the flow $\lambda_{ji}$ were not feasible, then choosing $t_i$ arbitrarily large or small would lead to a higher value for the Lagrangian, which yields a contradiction. Given that this is zero, the Lagrangian equals virtual welfare, and virtual welfare maximization follows from optimality.
\end{proof}

\subsection{A Sufficient Condition for Regular Distributions}

Our first set of sufficient conditions for upgrade pricing optimality consists of a weak monotonicity condition and a regularity condition. 

We call a type distribution $F$ \emph{weakly monotone} with cutoffs $i^1, i^2, \dots,i^d \in [n]$ if for any $i, j \in [n]$ and $k\in \{1, 2, \dots, d\}$,
\[
i \le i^k \le j \implies \theta_i^k \le \theta_j^k.
\]
Note that weak monotonicity is strictly weaker than monotonicity: for each item, only order comparisons  with respect to a cutoff type need to hold, whereas types above or below the cutoff can be arbitrarily ordered.

Similarly, a type distribution $F$ is \emph{regular} with respect to cutoffs $i^1, i^2, \dots,i^d \in [n]$ if for any $i, j \in [n]$ and $k\in \{1, 2, \dots, d\}$,
\begin{equation}
i \le i^k \le j \implies {\phi}_i^k \le 0 \le {\phi}_j^k, \label{eq:regular}
\end{equation}
where $\phi_i$ denotes the \emph{initial $d$-dimensional virtual values}
\begin{equation}
\phi_i \coloneqq \theta_i - \frac{1-F_{i}}{f_i}(\theta_{i+1} - \theta_i).
    \end{equation}
The initial $d$-dimensional virtual values can be seen as multi-dimensional versions of the virtual values in \cite{myer81}.

\begin{figure}[htbp]
\centering
\begin{subfigure}{.32\linewidth}
    \includegraphics[width=\linewidth]{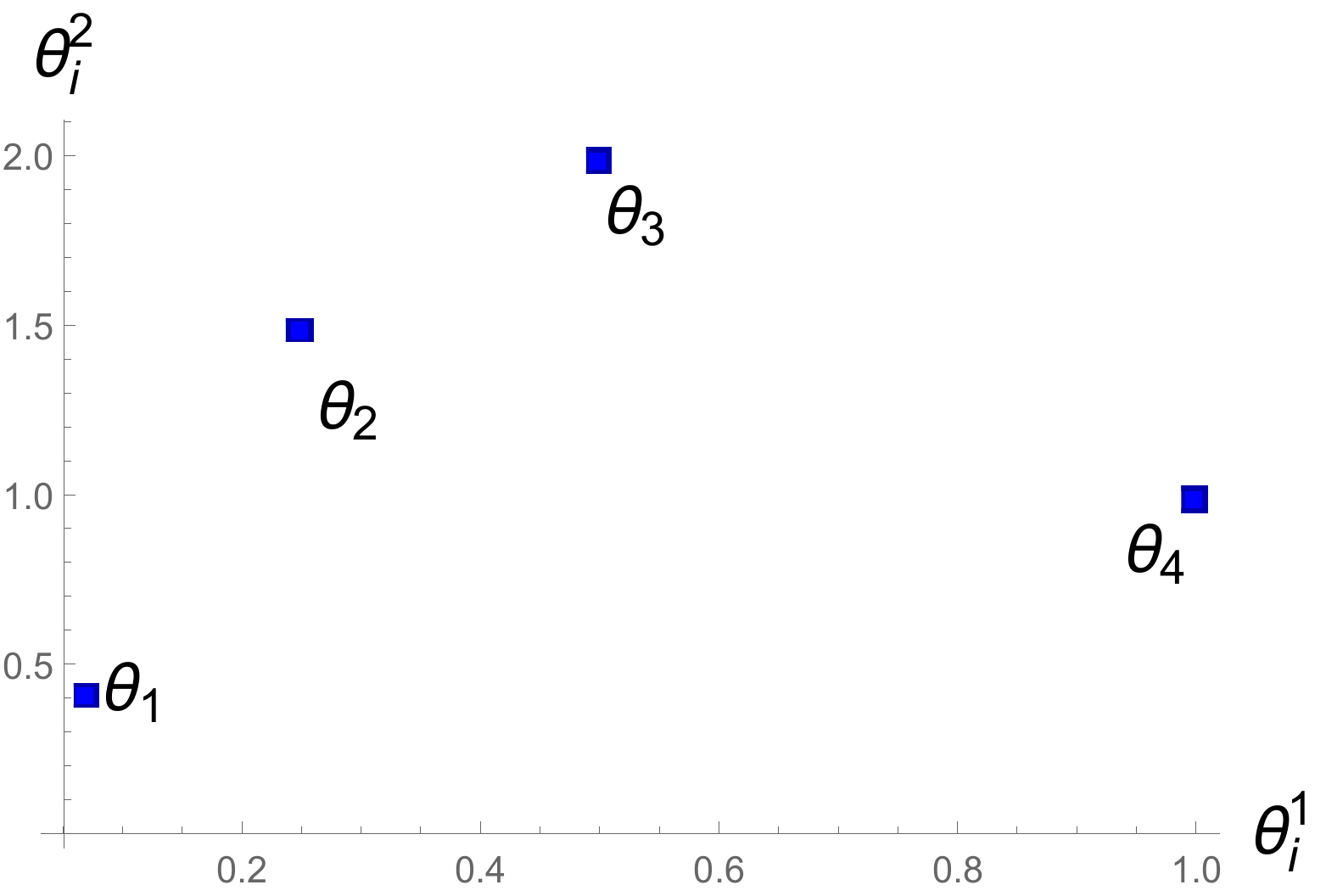}
    \caption{Types}
    \label{subfig:regtypes}
\end{subfigure}
\begin{subfigure}{.32\linewidth}
    \includegraphics[width=\linewidth]{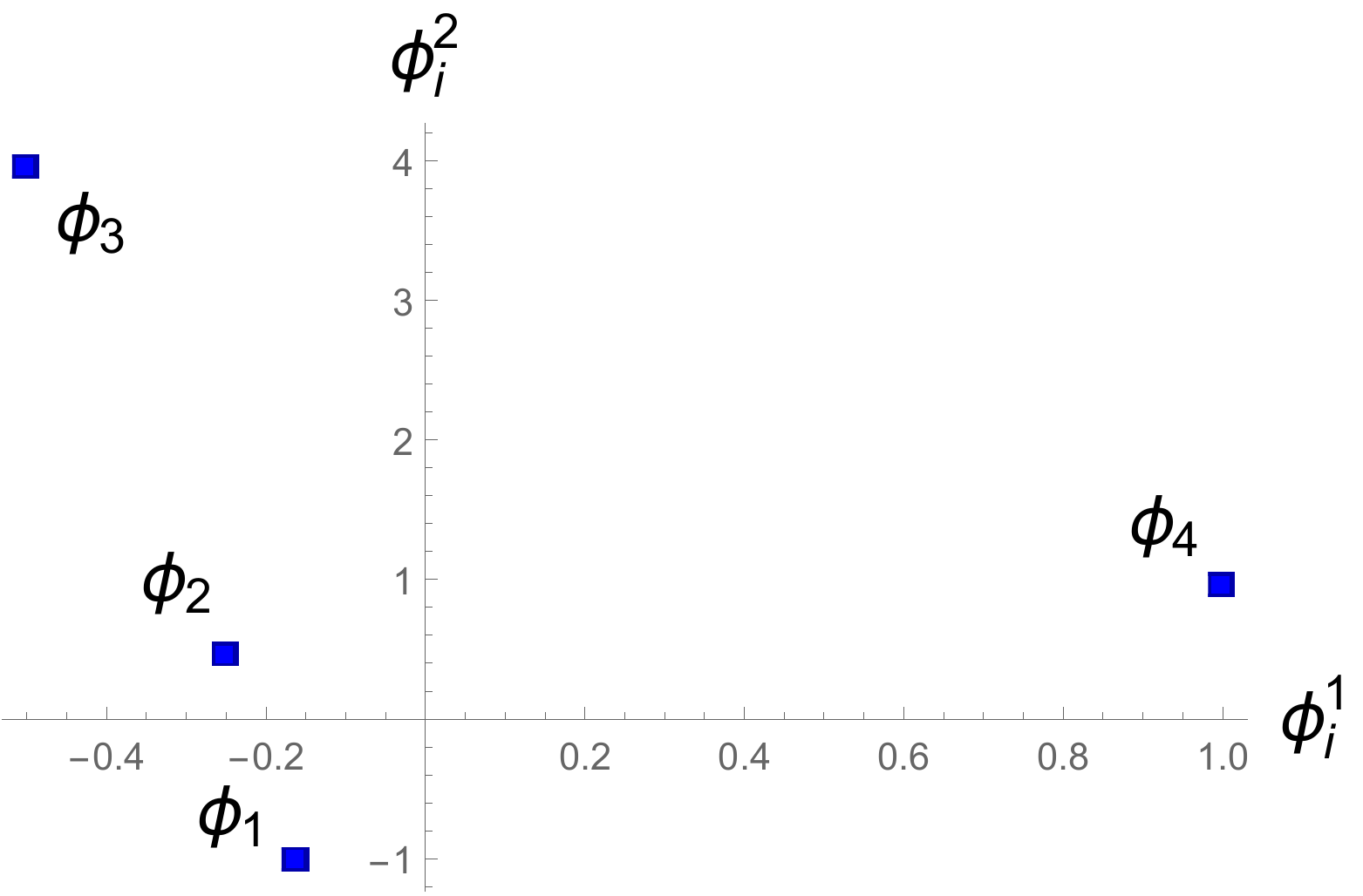}
    \caption{Virtual values}
\end{subfigure}
\begin{subfigure}{.32\linewidth}
    \includegraphics[width=\linewidth]{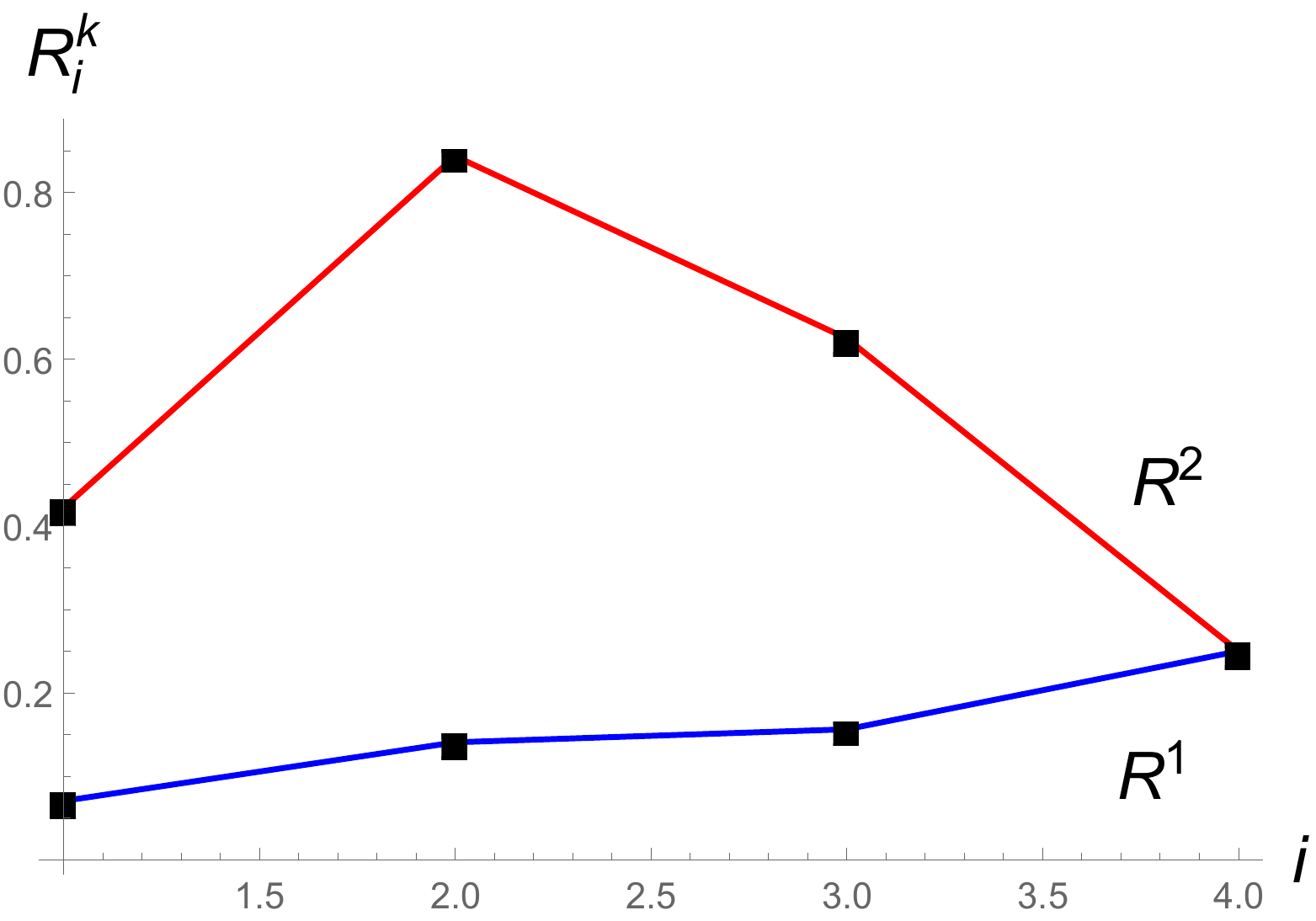}
    \caption{Pseudo-revenues}
    \label{subfig:pseudorev}
\end{subfigure}
      \label{fig:weak_monotone}
      \caption{Types, virtual values and pseudo-revenues for type space $\Theta=\{(\sfrac{9}{128},\sfrac{27}{64}),(\sfrac14,\sfrac32),(\sfrac12,2),(1,1)\}$ and type distribution $f=(\sfrac{7}{16},\sfrac{3}{16},\sfrac18, \sfrac14)$. The optimal mechanism sells good $2$ at a price of $1$ and good $1$ as an upgrade, also at a price of $1$. All types except $\theta_1$ buy good $2$, and only type $\theta_4$ buys good $1$.}
\end{figure}

We say that a type distribution $F$ is \emph{compatibly} weakly monotone and regular if it is both weakly monotone and regular with respect to the same set of cutoffs. When such cutoffs $i^k$ exist, they are essentially unique except between contiguous types of vanishing virtual value $\phi_i^k$ and monotone types $\theta_i^k$, $i \in [n]$, $k\in[d]$. Subfigure \ref{subfig:regtypes} illustrates a type distribution with this property.

Our regularity condition can be equivalently stated in terms of the \emph{pseudo-revenues}
\begin{equation}
 R_i^k \coloneqq (1-F_{i-1})\theta_i^k.\label{eq:pseudorevenue}
\end{equation}
Subfigure \ref{subfig:pseudorev} depicts pseudo-revenues. We call \eqref{eq:pseudorevenue} \emph{pseudo}-revenue because, without an assumption that the values are monotone with respect to the component-wise partial order, the pseudo-revenue does not correspond to the revenue from sales of item $k$ at a posted price of $\theta_i^k$. In particular, because we have
\begin{multline}
\frac{ R_i^k -  R_{i+1}^k}{f_i} = \frac{(1-F_{i-1}) \theta_i^k - (1-F_{i})\theta_{i+1}^k}{f_i} \\= \frac{f_i\theta_i^k - (1-F_i) (\theta_{i+1}^k - \theta_i^k)}{f_i} = \theta_i^k - \frac{1-F_i}{f_i} (\theta_{i+1}^k - \theta_i^k) = \phi_i^k   ,\label{eq:virtrev}
\end{multline}
imposing regularity with respect to the cutoffs $i^k$ is equivalent to requiring that  $R_i^k$ is single-peaked with peak $i^k$. While pseudo-revenues do not have immediate economic meaning, they are an important technical tool, in particular for our analysis of non-regular distributions in \autoref{sec:ironing}.

\begin{theorem}\label{thm:reg}
If the type distribution $F$ is compatibly weakly monotone and regular with respect to cutoffs $(i^k)_{k \in [d]}$, then upgrade pricing is optimal. In particular, the following mechanism is optimal:
\begin{equation}
\begin{split}
    q_i^k &\coloneqq \begin{cases} 1 &   i \ge i^k \\ 0 & \text{ else.} \end{cases}, \quad i \in [n], k \in [d].
\end{split}\label{eq:mechanism}
\end{equation}
\end{theorem}
\begin{proof}[Proof of \autoref{thm:reg}]
Define the dual variables
\begin{equation}\label{eq:initflow}
\hat \lambda_{ji} = \begin{cases}
1-F_{i} & \text{ if $j = i+1$} \\ 0 & \text{else.}
\end{cases}
\end{equation}
Observe that, by definition, $\hat \lambda$ induces the initial virtual values, $\phi_i = \phi_i^{\hat\lambda}$.

We check the properties of \autoref{lem:lagrange}. Virtual welfare maximization, condition \ref{enum:max}, follows from
\[
q_i^k = 1 \xLeftrightarrow{\text{\eqref{eq:mechanism}}} R_i^k \ge R_{i+1}^k \xLeftrightarrow{\text{\eqref{eq:virtrev}}} \phi_i^k \ge 0.
\]
For flow preservation, condition \ref{enum:flow}, observe that
\[
\sum_{j=1}^n \hat\lambda_{ij} - \sum_{j=0}^n \hat\lambda_{ji} = 1 - F_{i-1} - (1- F_i) = f_i.
\]
The mechanism is implementable, condition \ref{enum:feas}, by assumption of compatible weak monotonicity and regularity. 

Finally, we need to check that complementary slackness (condition \ref{enum:slack}) holds. Observe that $\hat\lambda_{ij} > 0$ implies $j = i-1$. Hence, all types must be indifferent between their allocation and payment and the allocation and payment of the next lower type. If the next lower type has the same allocation and payment, this is clearly satisfied. Otherwise, this is the first type buying an upgrade. If this type were not indifferent between buying it and not buying it, the price of the upgrade could be raised, and the revenue increased, without affecting other types' incentives. Thus,  this type must be indifferent between their allocation (and payment) and the next lower type's allocation.
\end{proof}

Our assumptions of regularity and weak monotonicity relax the monotonicity of types and Myersonian virtual values by allowing for permutations above and below the monopoly price. These assumptions nonetheless require that the set of types that buy each object remains an upper selection, and conversely the set of types that do not buy remains a lower selection. The intuition for why this works is similar to the idea that the monopoly price does not depend on the valuations of types that are not marginally buying, just as long as they do not become marginal buyers.

These assumptions depend on the fixed order of types we have introduced in the model. Thus, if there exists an order that satisfies these assumptions, upgrade pricing is optimal. Furthermore, multiple  orders of types might  satisfy the theorem's conditions for a given type distribution $F$. In this case, the theorem can be used to certify optimality of mechanism \eqref{eq:mechanism}, based on the different orders. As optimality of a mechanism for a distribution $F$ does not depend on the order on types, the revenue of \eqref{eq:mechanism} must be the same for all orders with which the conditions of \autoref{thm:reg} are satisfied.

Our next set of conditions imposes similar requirements, strengthened appropriately to allow for non-regular type distributions, which require ironing.

%% file: 4_ironing.tex
\section{Optimal Mechanisms for Non-Regular Distributions}\label{sec:ironing}
We now establish the optimality of an upgrade pricing mechanism in settings without regularity. The weaker sufficient conditions will replace the regularity condition and will allow for ironing to be part of the optimal mechanism.  The new sufficient conditions will serve to allow us to perform the ironing procedure item-by-item, and limit the interaction of constraints across items. 
We say that a type space $\Theta$ has \emph{monotone marginal rates of substitution} if
\[
1 \le i\le j \le n \text{ and } 1 \le k \le l \le d \implies  \frac{\theta_i^{l}}{\theta_i^k} \le \frac{\theta_j^{l}}{\theta_j^k}.
\]
for any $i, j \in [n]$, $l,k \in [d]$.

Recall that \emph{pseudo-revenue} is given by $R_i^k = (1-F_i)\theta_i^k$.

We call a scalar sequence $(R_i)_{i \in [n]}$, \emph{quasi-concave} if there is a cutoff $i'\in [n]$ such that $i' \le i \le j$ or $j \le i \le i'$ implies $R_i \ge R_j$. We call the point-wise smallest quasi-concave sequence that point-wise dominates $({R}_i)_{i \in [n]}$ its \emph{quasi-concave closure} and denote it by $(\overline{R}_i)_{i \in [n]}$. 

\begin{figure}[htbp]
    \centering
    \begin{subfigure}{.32\linewidth}
    \includegraphics[width=\linewidth]{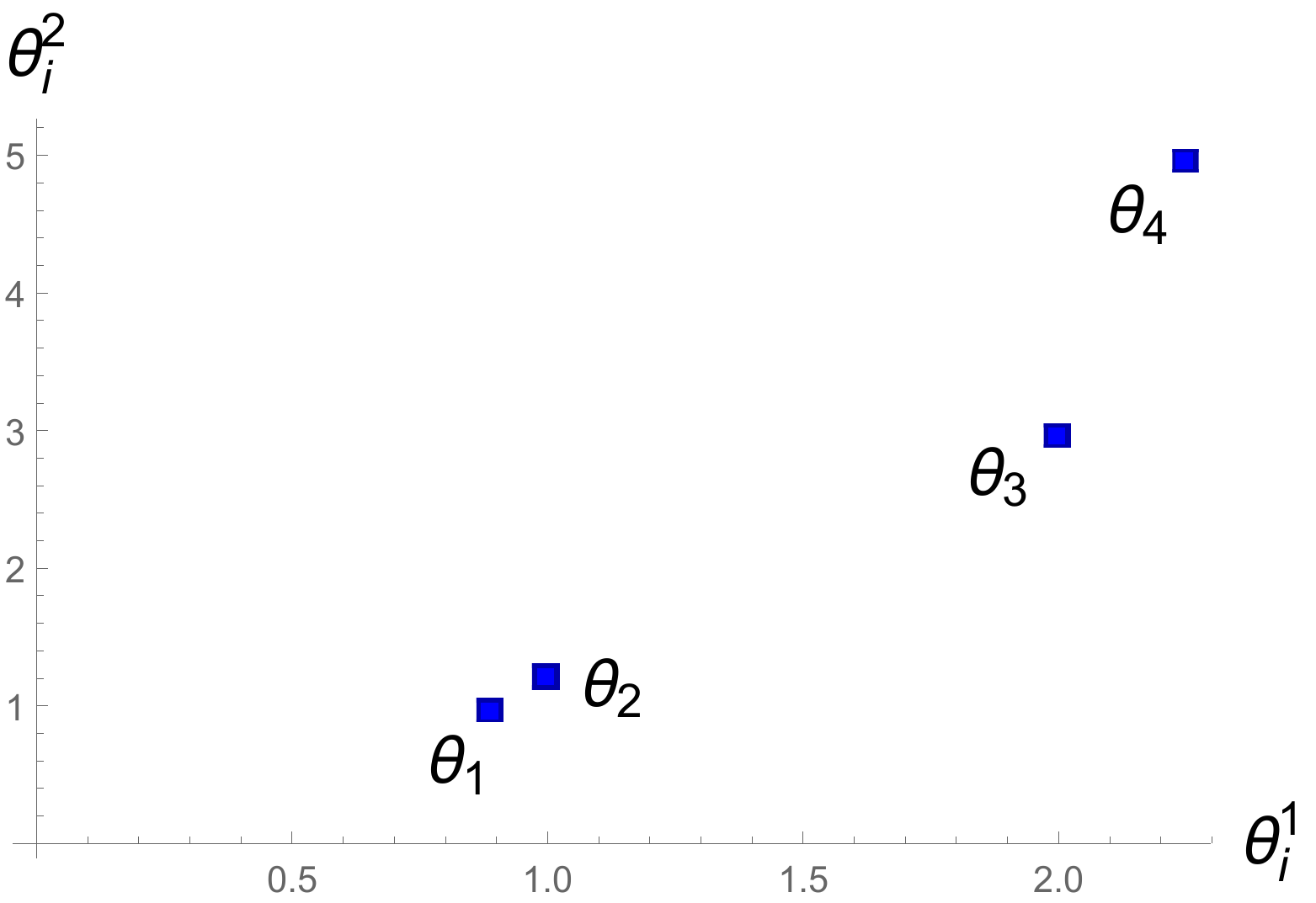}
    \caption{Types}
    \label{eq:nonregtypes}
  \end{subfigure}
\begin{subfigure}{.32\linewidth}
    \includegraphics[width=\linewidth]{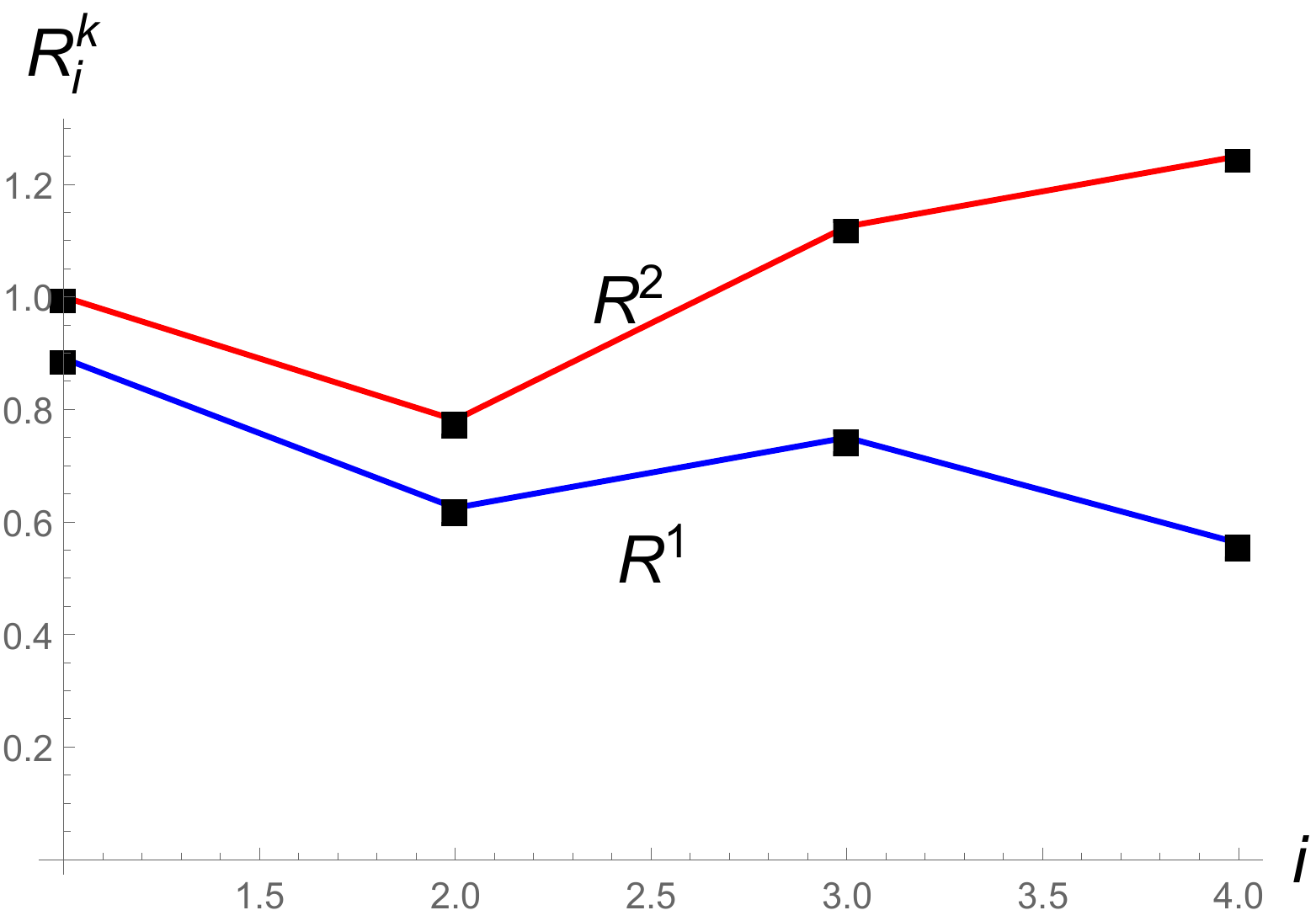}
    \caption{Pseudo-revenues}
    \label{subfig:pseudorev2}
    \end{subfigure}
    \caption{Type space and pseudo-revenues for type space $\Theta=\{(\sfrac{57}{64},1),(1,\sfrac54),(2,3),(\sfrac94,5)\}$ and type distribution $f=(\sfrac38,\sfrac14,\sfrac18,\sfrac14)$. The optimal mechanism sells good $1$ at a price of $\sfrac{57}{64}$, and good $2$ as an upgrade at a price of $5$. All types buy good $1$, and only type $\theta_4$ buys good $2$.}\label{fig:mrs}
\end{figure}

We will make regular use of the sequence $(\overline{R}^k_i)_{i \in [n]}$, the quasi-concave closure of the pseudo-revenue for item $k$.

To allow for our construction of a dual certificate of optimality, we need additional assumptions. These will be formulated in terms of  \emph{candidate ironing intervals}. For a pseudo-revenue $R$, we call a set of contiguous types $I \subseteq [n]$ with 
\begin{equation}
\overline{R}_i^k \neq R_i^k\label{eq:candironing}
\end{equation}
for all $i \in I$ such that there is no superset of contiguous types $I' \supseteq I$ such that \eqref{eq:candironing} holds for all $i \in I'$, a \emph{candidate ironing interval for item $k$}. (With slight abuse of language, we refer to discrete sets of contiguous types as \emph{intervals}.) Every item $k$ may have several candidate ironing intervals, and every type can be contained in a candidate ironing interval for different items.

We relax the regularity assumption on pseudo-revenues $R_i^k$. Instead of assuming regularity, i.e. $R_i^k$ to be single-peaked with peak $i^k$, we assume two properties that are in combination weaker than regularity.  We call a type distribution $F$ \emph{mostly regular} if for some cutoffs $i^k \in \argmax_{i \in [n]} R_i^k$ and any $i$ such that $i^k < i \le i^{k+1}$, the following hold:
\begin{enumerate}
    \item (No partial overlap) If $I$ is a candidate ironing interval of item $k$ and $J$ is a candidate ironing interval of item $k+1$, then either $I \cap J = \emptyset$ and there is $i \in [n]$ such that $I < i < J$ or $J < i < I$, or one of $I, J$ is a subset of the other excluding its endpoints.
    \item (No ironing on neighboring maxima) For any ironing candidate interval $I$ of item $k$, $i^k, i^{k+1}\notin I$. 
    \item (Not too shuffled) For any candidate ironing interval $I \subseteq \{i^k+1 , i^{k}+2, \dots, i^{k+1}-1\}$ and $i \in I$,
\begin{align*}
\theta_{\min I}^{k+1} &\le \theta_i^{k+1} & \theta_{\max I}^{k} &\le \theta_{\max I + 1}^{k} 
\end{align*}
\end{enumerate} 
Finally, we call a distribution \emph{compatibly} weakly monotone and mostly regular if it is weakly monotone and mostly regular with respect to the same cutoffs $i^k$, $k\in [d]$. 

Note that monotone MRS by itself is not a restrictive assumption. For example, in two dimensions, every type set can be ordered in order of monotone MRS. In combination with compatible weak monotonicity and mostly regularity, this assumption becomes stronger.

Subfigure \ref{eq:nonregtypes} shows the type space of a compatibly weakly monotone and mostly regular type distribution, and subfigure \ref{subfig:pseudorev2} its pseudo-revenues.

\begin{figure}[htbp]
   \centering
   \begin{subfigure}{.32\linewidth}
    \includegraphics[width=\linewidth]{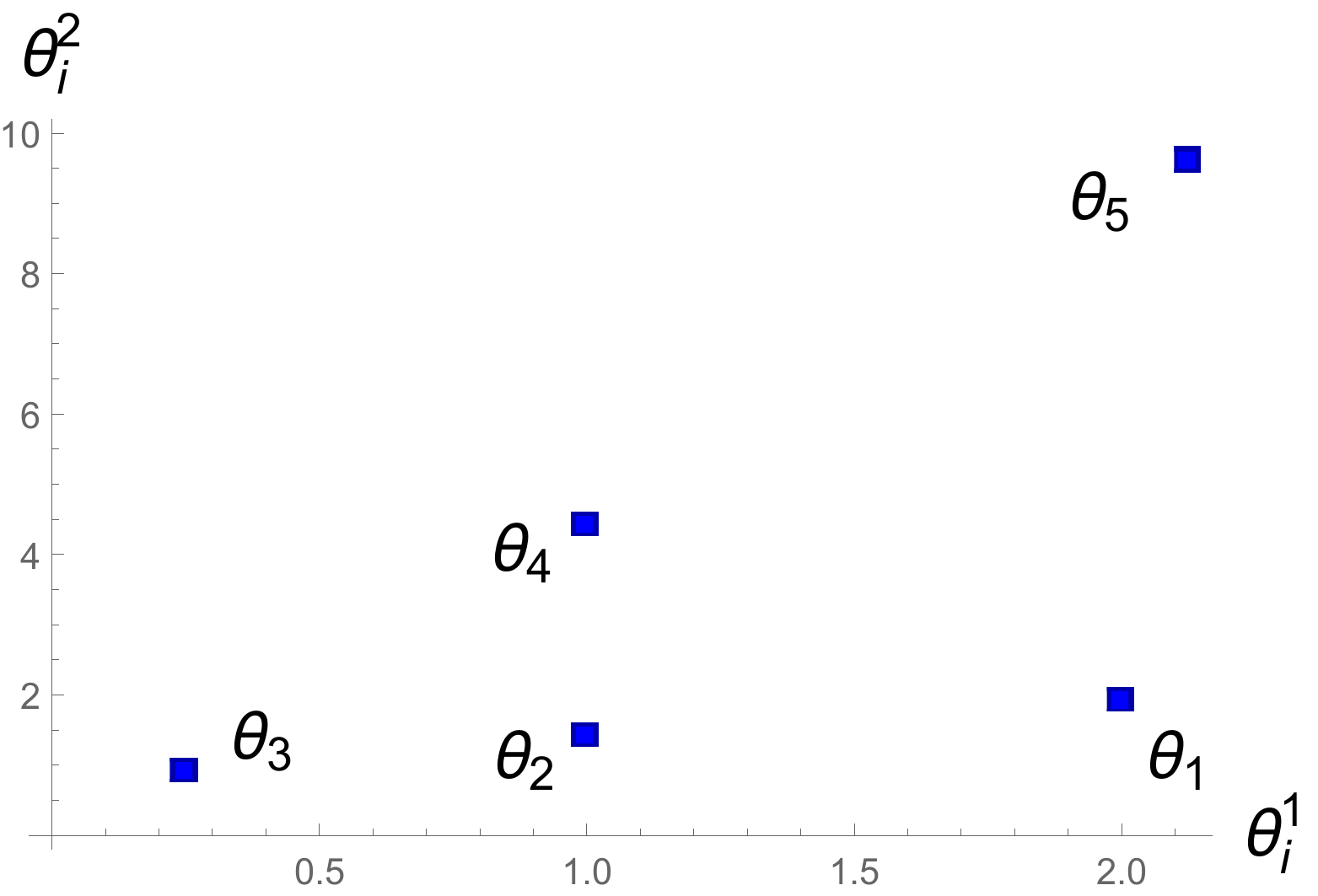}
    \caption{Types}
\end{subfigure}
\begin{subfigure}{.32\linewidth}
    \includegraphics[width=\linewidth]{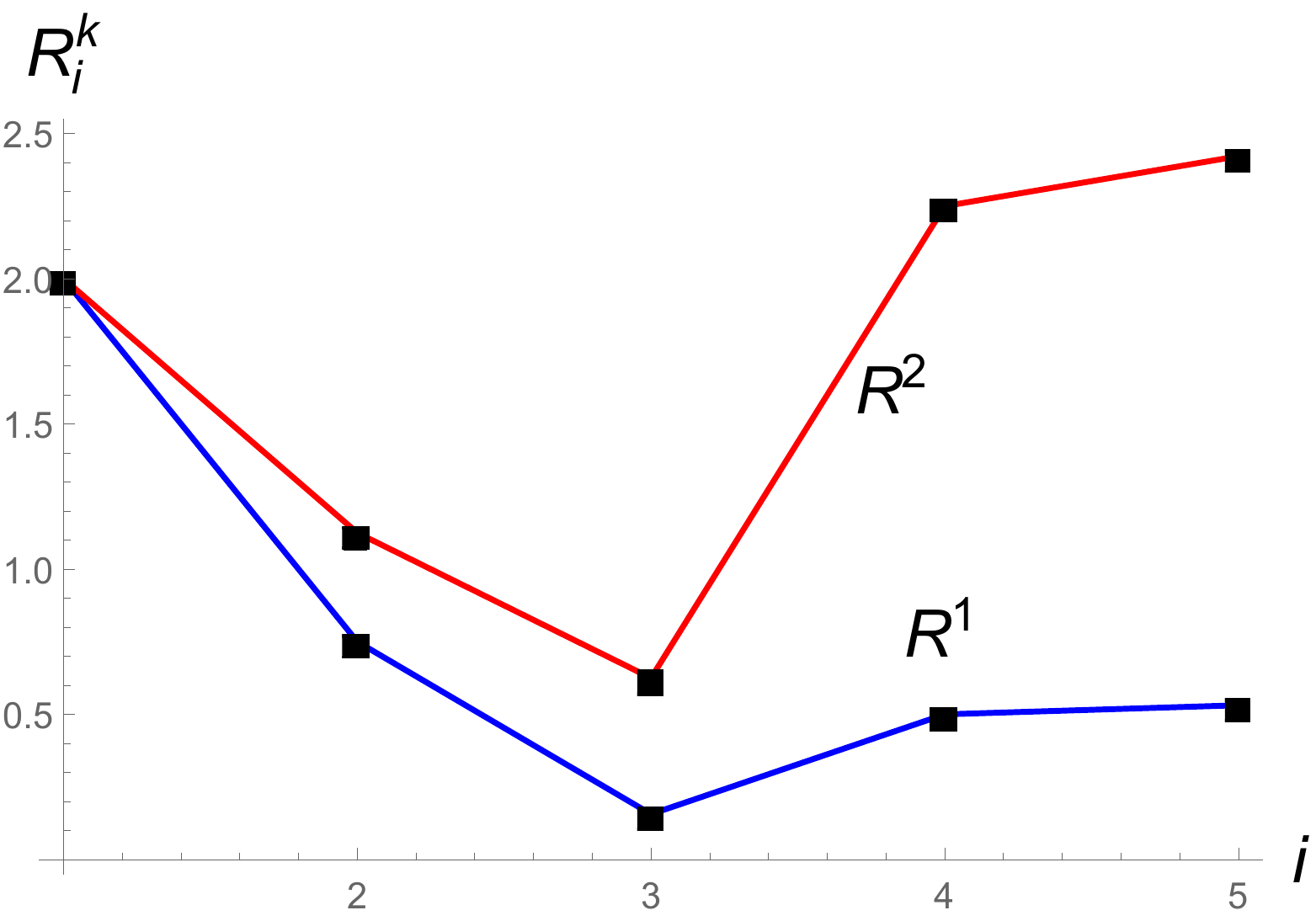}
    \caption{Pseudo-revenues}
\end{subfigure}
    \caption{Failure of no overlap: $\{2, 3\}$ is a candidate ironing interval for item 2, $\{3,4\}$ is a candidate ironing interval for item 1.}
    \label{fig:fail}
\end{figure}

Conversely, Figure \ref{fig:fail} shows an instance of a distribution over a monotone MRS type space that is \emph{not} mostly regular. In particular, this example fails the first condition, because it involves overlapping candidate ironing intervals. 
We will use our assumptions to construct dual variables $(\lambda_{ij})_{i,j \in [n]}$ by ironing pseudo-revenues for each item. In our proof that there is an optimal mechanism with an upgrade pricing allocation, we will use monotone MRS to show that for each type, ironing is only needed for two items, the lowest item in the MRS order that the type bought, and the highest item in the MRS order that she didn't buy. We will use the first two conditions of mostly regularity to show that from these two items, we can select a single item to iron at a time, while not changing the other item's virtual values in a way that will break virtual welfare maximization of the allocation. As in Theorem \ref{thm:reg}, weak monotonicity ensures implementability of an upgrade pricing allocation, i.e., the existence of a price vector $(t_i)_{i \in [n]}$ such that the mechanism $(q, t)$ is incentive compatible and individually rational. To allow for our ironing procedure to work, we also need a mild requirement on the monotonicity of types beyond weak monotonicity. While weak monotonicity was a requirement that could be formulated item-by-item, this requirement links the type order of neighboring items. 
\begin{theorem}\label{thm:mrs}
Let $\Theta$ have monotone marginal rates of substitution. If the type distribution $F$ is compatibly weakly monotone and mostly regular with respect to cutoffs $(i^k)_{k \in [d]}$, then upgrade pricing is optimal. In particular, the following mechanism is optimal:
\begin{equation}
    q_i^k \coloneqq \begin{cases} 1 &  i \ge i^k \\ 0 & \text{else,}\end{cases}\quad i \in [n], k \in [d].
\label{eq:mechanism2}
\end{equation}
\end{theorem}
Note that the allocation \eqref{eq:mechanism2} is the allocation that arises from separate monopoly pricing.\footnote{In \autoref{sec:sep}, we further explore the relationship between upgrade pricing and separate pricing, by showing conditions under which the allocation \eqref{eq:mechanism2} can be implemented by a vector of single-item prices.}

\begin{figure}
    \centering
    \begin{subfigure}{.32\linewidth}
    \includegraphics[width=\linewidth]{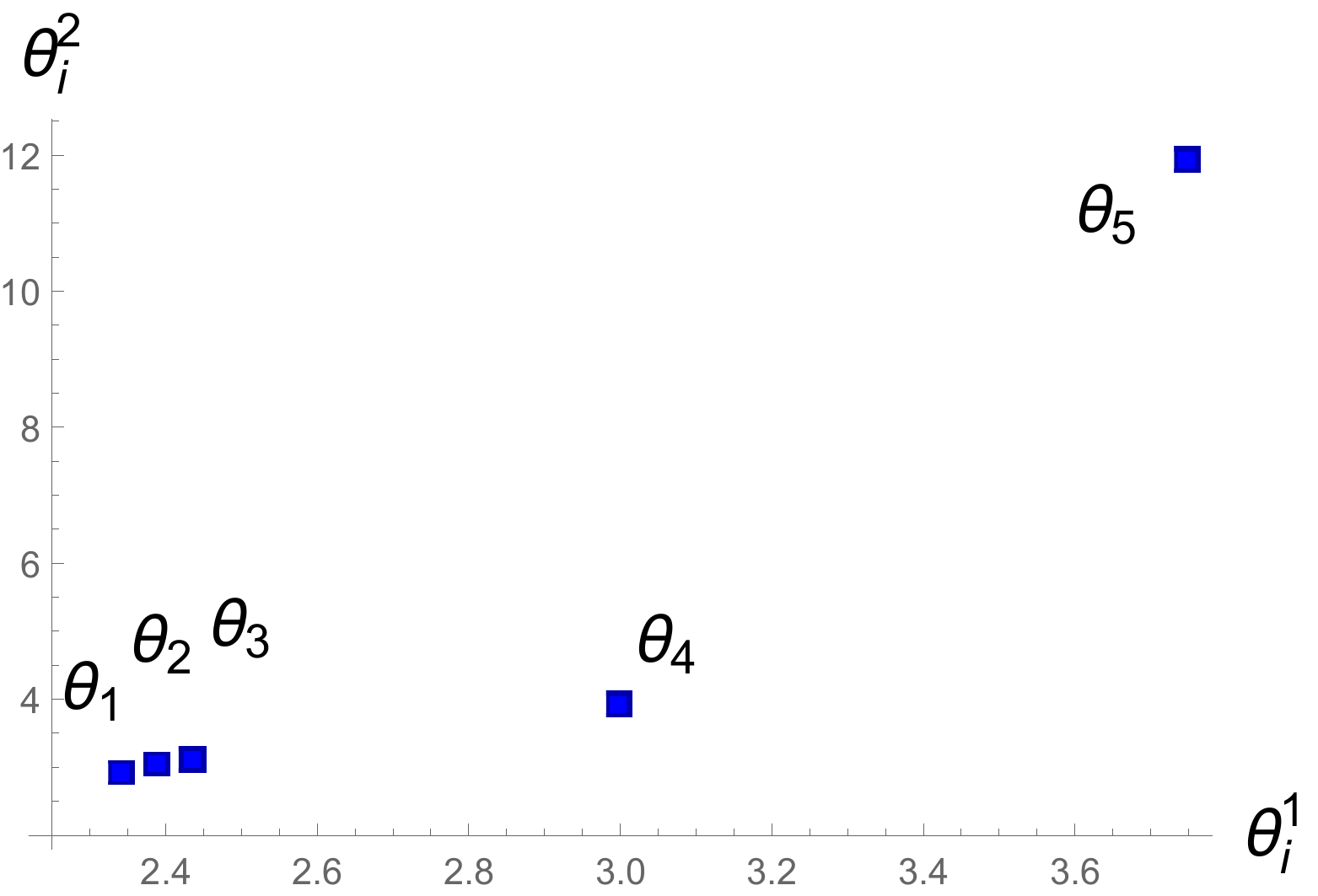}
    \caption{Types}
  \end{subfigure}
  \begin{subfigure}{.32\linewidth}
    \includegraphics[width=\linewidth]{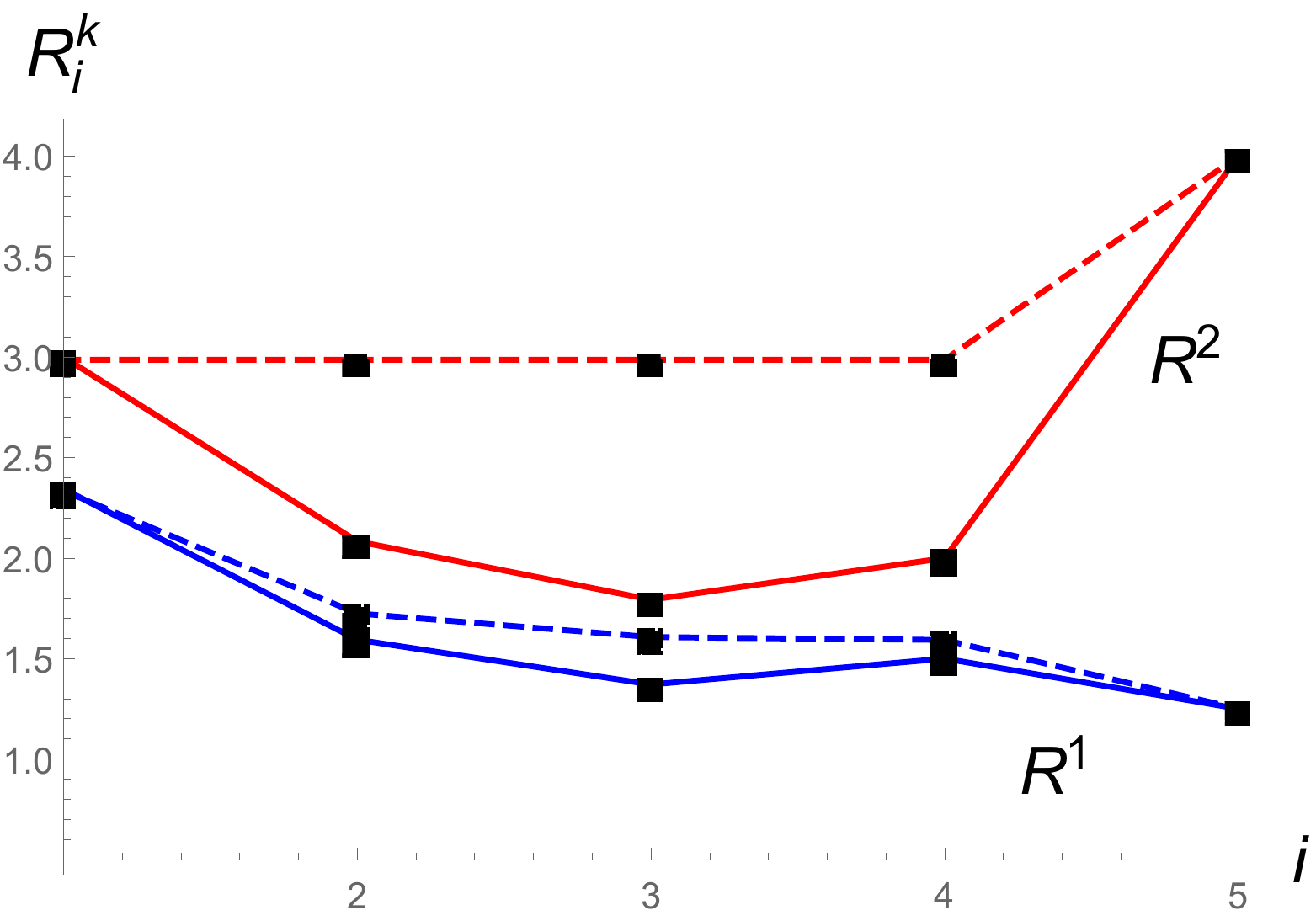}
    \caption{Pseudo-revenues, ironed pseudo-revenues (dashed)}
  \end{subfigure}
    \caption{Ironing of virtual values and corresponding pseudo-revenues.}
    \label{fig:firststeps}
\end{figure}

To prove \autoref{thm:mrs}, we will construct a sequence of flows $\lambda_i$ from $i=n$ down to $i=1$, starting with $\hat \lambda$, the initial flow that induces Myersonian multi-dimensional virtual values. Given a definition of pseudo-revenue implied by a flow, our Ironing Algorithm will, for each type $i$ and for at least one item $k$, iron to match induced pseudo-revenue with the quasi-concave closure of multi-dimensional Myersonian pseudo-revenue, $\overline{R}^k_i$. This is illustrated in \autoref{fig:firststeps}. 

The main steps in this proof are to show that the ironing is well-defined in that such implied pseudo-revenue is attainable with a non-negative and feasible flow (\autoref{lem:algowelldef} and \autoref{lem:feasnonneg}, respectively). The most technical part of the proof consists of showing that the Ironing Algorithm produces dual variables that maximize virtual welfare (\autoref{lem:weakvirtwelfmax} and \autoref{lem:vwfcsl} (a)), and satisfy complementary slackness (\autoref{lem:vwfcsl} (b)). 

Our first lemma is a main structural tool to link different items' virtual values and is tightly connected to monotone MRS. For $k \in [d]$, $i \in [n]$, and  flow $\lambda$, denote the normalized virtual value by
$$
\nu^{\lambda,k}_{i} \coloneqq  \frac{\phi_i^{k,\lambda}}{\theta_i^{k}}.$$
The property that we will use repeatedly is that $\nu_i^{\lambda,k}$ has the same sign as $\phi_i^{\lambda, k}$. We call a flow \emph{downward} if $\lambda_{ji} > 0$ for $i,j \in [n]$ implies that $j > i$.
\begin{lemma}\label{lem:localordered}
Let $\Theta$ have monotone MRS. For any non-negative downward flow $\lambda$, $\nu^{\lambda,k}_{i} \ge \nu^{\lambda,l}_{i}$ for any $1 \le k \le l \le d$ and $i \in [n]$.
\end{lemma}
\begin{proof}
It follows from definitions and monotone marginal rates of substitution that
\begin{align*}
    \frac{\phi^{\lambda,k}_i}{\theta_i^k} &= \frac{\theta_i^k - \frac{1}{f_i}\sum_{j=1}^n \lambda_{ji} (\theta_j^k - \theta_i^k)}{\theta_i^k} = 1 + \frac{1}{f_i}\sum_{j=i}^n \lambda_{ji} - \frac{1}{f_i}\sum_{j=i}^n \lambda_{ji} \frac{\theta_j^k}{\theta_i^k} \\
    & \ge 1 + \frac{1}{f_i}\sum_{j=i}^n \lambda_{ji} - \frac{1}{f_i}\sum_{j=i}^n \lambda_{ji} \frac{\theta_j^l}{\theta_i^l} = \frac{\theta_i^l - \frac{1}{f_i}\sum_{j=1}^n \lambda_{ji} (\theta_j^l - \theta_i^l)}{\theta_i^l}=\frac{\phi^{\lambda,l}_i}{\theta_i^l}.
\end{align*}
\end{proof}
The next Lemma shows that virtual welfare maximization reduces to virtual welfare maximization for the neighboring items, i.e., the last item that a type buys and the first item that a type does not buy---with respect to the MRS order.
\begin{lemma}\label{lem:weakvirtwelfmax}
Assume  $\Theta$ has monotone MRS and mostly regular and that there exists a non-negative downward flow $\lambda$ such that for any $i \in [n]$ such that $i^k \le i \le i^{k+1}$, we have $\phi_i^{\lambda,k}  \ge 0$ and $\phi_i^{\lambda,k+1} \le 0$. 
Then, the allocation in \eqref{eq:mechanism2} maximizes virtual welfare.
\end{lemma}
\begin{proof}
Fix $i \in [n]$ such that $i^k \le i \le i^{k+1}$. Note that as $\phi_i^{\lambda, k}$ and $\nu_i^{\lambda, k}$ are positive multiples of each other, \autoref{lem:localordered} implies the implications 
\begin{align*}
\phi_i^{\lambda, k+1} \le 0 &\implies \phi_i^{\lambda, l} \le 0, \qquad l \ge k+1\\
\phi_i^{\lambda, k} \ge 0 &\implies \phi_i^{\lambda, l} \ge 0, \qquad l < k.
\end{align*}
Therefore, the assumption implies that
$\phi_i^{\lambda, l} \le 0$ for any $l > k$ and $\phi_i^{\lambda, l} \ge 0$ for any $l\le k$, which ensures virtual welfare maximization of \eqref{eq:mechanism2}.
\end{proof}
For $k = 0$ and $k=d$ this Lemma reduces virtual welfare maximization for all items, and ironing for all items, to virtual welfare maximization for the first resp. last item. Finding a flow that maximizes virtual welfare reduces to ironing the (one-dimensional) virtual values $\phi_i^1$ and $\phi_i^d$. For types $i \le i^1$ and $i \ge i^d$, we can hence use techniques from one-dimensional ironing and iron the pseudo-revenue to its concave closure in a discrete variant of \cite{myer81}'s procedure. From now, our discussion therefore focuses on $k \in [d-1]$ and $i \in [i^k+1, i^{k+1}]$, i.e. types where an ironing that ensures virtual welfare maximization for both item $k$ and item $k+1$ is needed.

The following algorithm will make use of $\hat \lambda$ as defined in \eqref{eq:initflow}, the \emph{initial flow} and of a generalization of the pseudo-revenue. The pseudo-revenue associated to a flow $\lambda$, $R_i^{\lambda, k}$ is
\[
R_i^{\lambda,k} = \sum_{j=i}^n f_j \phi_j^{\lambda,k}.
\]
This generalization is intuitive, as virtual values are, as in \eqref{eq:virtrev}, slopes of pseudo-revenues
\begin{equation}
\frac{R_{i}^{\lambda, k} -R_{i+1}^{\lambda, k}}{f_i}  = \frac{\sum_{j=i}^n f_j \phi_j^{\lambda,k} -\sum_{j=i+1}^n f_j \phi_j^{\lambda,k}}{f_{i}} =  \phi_i^{\lambda,k}.\label{eq:virtslopes}
\end{equation}
Our algorithm will adjust a flow by raising one point in a revenue sequence at a time, from right to left. We will prove that this will yield slopes of revenue sequences---i.e. virtual values---which have the correct sign for virtual welfare maximization of \eqref{eq:mechanism2}. This is non-trivial, as pseudo-revenues for different items might not move in the same direction when dual variables are changed.
\begin{algorithm}[H]
\SetAlgoVlined
$\lambda \leftarrow \hat{\lambda}$\;
\For{$i=n$ \KwTo $1$}{
Let $\gamma_i \in [0,1]$ be maximal such that for
        \begin{equation}
        \begin{split}
        \lambda'_{ji} &\leftarrow \gamma_i \lambda_{ji}, \quad \forall  j: n > j > i\\
    \lambda'_{j(i-1)} &\leftarrow \lambda_{j(i-1)} + (1-\gamma_i) \lambda_{ji}, \quad \forall j: n > j > i\\
    \lambda'_{i(i-1)} &\leftarrow \lambda_{i(i-1)} - (1-\gamma_i) \sum_{i'=i}^{n} \lambda_{i'i},
    \end{split}\label{eq:parameterizedflow}
        \end{equation}
     $R_i^{\lambda', \kappa(i)} = \overline{R}^{\kappa(i)}_i$ holds\;
    $\lambda \leftarrow \lambda'$\;
    }
Return $\lambda'$\;
\caption{Ironing, parameterized by an ironing mapping $\kappa\colon[n]\to[d]$}\label{algo:ironing}
\end{algorithm}
The flow \eqref{eq:parameterizedflow} was used earlier in \cite{Haghpanah2020}. An important difference is that \cite{Haghpanah2020} choose $\gamma_i$ to iron the revenue sequence of the grand bundle to the concave closure of pseudo-revenue. Instead, we iron to the \emph{quasi}-concave closure of (their equivalent of) pseudo-revenue of an item $\kappa(i)$. The parameter $\gamma_i$ can be found as solution to a system of linear equations. We show that a solution $\gamma_i \in [0,1]$ exists in \autoref{lem:algowelldef}.

We first observe that the Ironing Algorithm outputs a flow which is non-negative and feasible.
\begin{lemma}\label{lem:feasnonneg}
The output of the Ironing Algorithm is a flow, i.e. non-negative and satisfies flow preservation, \autoref{lem:lagrange} \autoref{enum:flow}.
\end{lemma}
\begin{proof}
We prove the claim by induction from $i = n$ to $i=1$. Note that $\hat\lambda$ is feasible as argued in the proof of \autoref{thm:reg}, which starts the induction. Fix an arbitrary iteration $i \in [n]$, and assume that $\lambda$ is feasible. We check that the difference in excess flow, i.e. incoming and outgoing flow, cancel out for $j>i$. For $i$ and $i+1$ similar calculations yield the result. We omit these. For any $j > i$, 
\[
\lambda'_{ji} - \lambda_{ji} + \lambda'_{j(i-1)} - \lambda_{j(i-1)} = \gamma\lambda_{ji} - \lambda_{ji} + \lambda_{j(i-1)} + (1-\gamma_i)\lambda_{ji}- \lambda_{j(i-1)} = 0.
\]
Now consider non-negativity. Each $\lambda_{ij}$ reduces at most once during the course of the Ironing Algorithm. More specifically, only if $j = i-1$ and during iteration $i$. In this iteration, 
\[
\lambda'_{i(i-1)} = \lambda_{i(i-1)} - (1-\gamma_i) \sum_{i'=i}^{n} \lambda_{i'i} \ge \lambda_{i(i-1)} - \sum_{i'=i}^{n} \lambda_{i'i} = f_i \ge 0,
\]
where we used that $\gamma_i \le 1$,  $\lambda_{ir} =  \hat\lambda_{ir}$, $r < i-1$, which in particular implies that $\lambda_{ir} = 0$, and feasibility of the flow.
\end{proof}

Next observe that in the Ironing Algorithm, iteration $i$ changes the revenue (for any item $k$) only for type $i$. Hence, our ironing algorithm raises pseudo-revenue for one type at a time.
\begin{lemma}\label{lem:changeslocal}
For any iteration $i$, $R_{j}^{\lambda',k} = R_{j}^{\lambda,k}$ for any $j \neq i$. In particular, $\phi_{j}^{\lambda',k} = \phi_{j}^{\lambda,k}$ for $j \notin \{ i-1, i\}$. 
\end{lemma}
\begin{proof}
First note that as the in-flow for higher types remains unchanged in iteration $i$, $\lambda'_{rj} = \lambda_{rj}$, $r \in [n]$, $j > i$, the revenue does not change, $R_j^{\lambda',k} = R_j^{\lambda,k}$. For types $j < i$, we check that the changes to virtual welfare on the types whose inflows do change, $i$ and $i-1$, cancel out. By definition of virtual values,
\begin{align*}
    \phi^{\lambda',k}_{i} &= \phi^{\lambda,k}_{i} + \frac{1-\gamma_i}{f_{i}} \sum_{j=i}^{n} \lambda_{ji} (\theta_j^{k} - \theta_{i-1}^{k})  \\ 
    \phi^{\lambda',k}_{i-1} &= \phi^{\lambda,k}_{i-1}  + \frac{1-\gamma_i}{f_{i-1}} \sum_{j=i}^{n} \lambda_{ji} (\theta_j^{k} - \theta_{i-1}^{k}) - \frac{1-\gamma_i}{f_{i-1}} \sum_{j=i}^{n} \lambda_{ji} (\theta_j^{k} - \theta_{i-1}^{k})\\
    &= \phi^{\lambda,k}_{i} - \frac{1-\gamma_i}{f_{i}} \sum_{j=i}^{n} \lambda_{ji} (\theta_j^{k} - \theta_{i-1}^{k}).
\end{align*}
Hence, 
\[
f_{i-1} \phi^{\lambda', k}_{i-1} + f_{i} \phi^{\lambda',k}_{i} = f_{i-1} \phi^{\lambda,k}_{i-1} + f_{i} \phi^{\lambda,k}_{i}.
\]
The statement on the virtual values follows from \eqref{eq:virtslopes}.
\end{proof}
Before showing that $\gamma_i$ in the algorithm always exists, we define the ironing function $\kappa(i)$. 

By no ironing on neighboring maxima, each candidate ironing interval $I$ must be contained in an interval $\{i^k, i^{k}+1, \dots, i^{k+1}\}$. By this condition, in addition to no partial overlap, for each type $i$, there is a unique inclusion maximal candidate among the candidate ironing intervals for items $k$ and $k+1$. We let $\kappa(i)$ denote the item this interval is a candidate ironing interval for. If $i$ is not part of any ironing interval, we set $\kappa(i)$ arbitrarily in $\{k, k+1\}$. We call $\kappa(i)$ the \emph{ironed item for type $i$} and piece-wise constant intervals of $\kappa$ \emph{ironing intervals}.
\begin{lemma}\label{lem:algowelldef}
Assume that $F$ is mostly regular. Then, for each $i \in [n]$, $\gamma_i$ such that $R_i^{\lambda_i(\gamma_i), \kappa(i)} = \overline{R}^{\kappa(i)}_i$ exists. In particular, the Ironing Algorithm is well-defined.
\end{lemma}
\begin{proof}
We prove this statement by induction from $i=n$ down to 1. Let $i \in [n]$ and assume that $R_{i+1}^{\lambda, \kappa(i)} = \overline{R}_{i+1}^{\lambda, \kappa(i)}$.  If $i$ is not part of an ironing interval, then by definition of ironing intervals and \autoref{lem:changeslocal}, $R_i^{\lambda, k} = \overline R_i^{\lambda, k}$, and the induction step is trivial by choosing $\gamma_i = 1$, yielding $R_i^{\lambda'(1), k} = \overline R_i^{\lambda'(1), k}$. Otherwise, $i$ is in an ironing interval. Let $\kappa(i) = k$.  By no partial overlap, if $i+1$ is part of an ironing interval, it must be part of the same ironing interval, in particular must have been ironed for item $k$. Hence, by the induction hypothesis, $R_{i+1}^{\lambda, k} = \overline{R}_{i+1}^{\lambda, k}$. 

Denote 
\[
\overline{{\phi}}^k_i = \frac{\overline{R}^k_{i} - \overline{R}^k_{i+1}}{f_i}
\]
the slope of the quasi-concave closure of pseudo-revenue of item $k$ at type $i$. By definition of the quasi-concave closure, the slope of the revenue curve must be non-positive, \[
\overline{{\phi}}^k_i \le 0.
\]
As all types are non-negative, we get that 
\begin{equation}
    \overline{{\phi}^k_i} \le 0 \le \theta_i^k = \phi_i^{\lambda'(0),k}.\label{eq:lowerboundvirt}
\end{equation}

Again by \autoref{lem:changeslocal}, $\overline{R}^k_{i+1} = R^{\lambda'(0),k}_{i+1}$. Therefore
\begin{align*}
\overline{R}^k_i &= f_i\overline{{\phi}^k_i}  + \overline{R}^k_{i+1} =f_i\overline{{\phi}^k_i}  + {R}^{\lambda'(0),k}_{i+1}\\
&\le f_i\phi^{\lambda'(0),k}_i  + {R}^{\lambda'(0),k}_{i+1} = {R}^{\lambda_i(0),k}_{i}.
\end{align*}
In particular, $\overline{R}^k_i \le {R}^{\lambda_i(0),k}_{i}$.

Also, by \autoref{lem:changeslocal} and the definition of the quasi-concave closure, $R_i^{\lambda'(1),k} = R_i^{\lambda,k} \le \overline{R}^k_i$. As $\gamma \mapsto R_i^{\lambda'(\gamma),k}$ is a continuous function, the existence of the desired $\gamma \in [0,1]$ follows from the Intermediate Value Theorem.
\end{proof}
The last lemma before the proof of Theorem \ref{thm:mrs} shows that the output of the algorithm satisfies complementary slackness and the condition of \autoref{lem:weakvirtwelfmax}, which is sufficient for virtual welfare maximization.
\begin{lemma}\label{lem:vwfcsl}
Assume that $\Theta$ is has monotone MRS, and that $F$ is mostly regular. Then, $q$ maximizes virtual welfare and satisfies the requirements of \autoref{lem:weakvirtwelfmax} with respect to $\lambda'$, the output of the Ironing Algorithm.
\end{lemma}
\begin{proof}[Proof of \autoref{lem:vwfcsl}]
We first show that $\lambda'$ satisfies the requirements of \autoref{lem:vwfcsl}, i.e. that for any $i^k \le i \le i^{k+1}$ we have that $\phi_i^{\lambda,k} \ge 0$ and $\phi_i^{k+1, \lambda} \le 0$. Let $i \in [n]$. If $i$ is not in an ironing interval, and there is no ironing interval $I$ such that $i = \min I-1$, the claim follows from the definition of the quasi-concave closure and our definition of ironing intervals, as well as \autoref{lem:changeslocal}.

We consider the remaining cases $i = \max I$, $i = \min I -1$, and $i \in I \setminus \{\max I\}$ separately. For the third case, by definition of the ironing and the quasi-concave closure,
\[
\phi_i^{\lambda',\kappa(i)} = \nu_i^{\lambda',\kappa(i)} =0.
\]
For $\kappa(i) = k$, $\nu_i^{k+1, \lambda} \le 0$ by \autoref{lem:localordered} and hence $\phi_i^{k+1, \lambda} \le 0$ as $\nu_i^{k+1, \lambda}$ and $\phi_i^{k+1, \lambda}$ are positive multiples of each other. Similarly, we have for $\kappa(i) = k+1$ that $\nu_i^{k, \lambda} \ge 0$ by \autoref{lem:localordered} and hence $\phi_i^{k, \lambda} \ge 0$. 
It remains to consider $i = \min I -1$ and $i = \max I$. We consider these separately for $\kappa(i) =k$ and $\kappa(i) = k+1$, for a total of four cases.

Case 1. $i = \max I$ and $\kappa(i) = k$. By definition of the quasi-concave closure, $\overline{R}^{k}_i =  R_i^{k}$, the algorithm chooses $\gamma_i = 1$ and hence $\phi_i^{\lambda', k} = \phi_i^{k}  = 0$ and hence by arguments as above, $\phi_i^{\lambda', k+1} = \phi_i^{k+1}  \le 0$.

Case 2. $i=\min I-1$ and $\kappa(i) = k+1$. By definition of the quasi-concave closure, $\overline{ R^{k+1}_i} =  R_i^{k+1}$, the algorithm chooses $\gamma_i = 1$ and hence, by an argument similar to case 1, $\phi_i^{\lambda, k} = \phi_i^{k}  \ge 0$.

Case 3. $i = \max I$ and $\kappa(i) = k+1$. 

The derivative of the virtual value of the right end of the ironing interval, type $i$ is given by
\begin{align}
  \frac{\partial \phi_{i}^{\lambda',k}}{\partial \gamma} &= -\frac{1}{f_i}\sum_{j =i+1}^n \lambda_{ji} (\theta^{k}_j -\theta^{k}_i)= -\frac{1}{f_i} \lambda_{(i+1)i} (\theta^{k}_{i+1} -\theta^{k}_i).\label{eq:case3nonmon}
\end{align}

By no partial overlap, $0 \le  \phi_i^{k}$. Also, by not too shuffledness, \eqref{eq:case3nonmon} is non-positive. Hence, $ \phi_i^{k} =\phi_i^{\gamma'(0),k}\le \phi_i^{\lambda', k}$, and the algorithm chooses $\gamma_i < 1$. Combining these observations, we obtain $\phi_i^{k, \lambda'} \ge 0$.

Because the algorithm chooses $\gamma_i < 1$, this implies 
\[
\phi_i^{\lambda', k} \ge \phi^{k}_i  \ge 0.
\]

Case 4. $i=\min I-1$ and $\kappa(i) = k$.
The derivative of the virtual value of the next item at the left end of the ironing interval, type $i$ is given by
\begin{equation}
\begin{split}
  \frac{\partial \phi_{i}^{\lambda',k+1}}{\partial \gamma} &= \frac{1}{f_i}\sum_{j =i+1}^n \lambda_{ji} (\theta^{k+1}_j -\theta^{k+1}_i) - \frac{1}{f_i}\sum_{j =i+1}^n \lambda_{ji} (\theta^{k+1}_{i+1} -\theta^{k+1}_i)\\
  &= \frac{1}{f_i}\sum_{j =i+1}^n \lambda_{ji} (\theta^{k+1}_j -\theta^{k+1}_{i+1})\ge 0,
\end{split}\label{eq:nonregcase4}
\end{equation}
where the last inequality follows from not-too-shuffledness (note that $i+1 = \min I$). By no partial overlap, $0 \ge  \phi_i^{k+1}$. Moreover, $ \phi_i^{k+1} =\phi_i^{\gamma'(0),k}\ge \phi_i^{\lambda', k+1}$ because of \eqref{eq:nonregcase4} and because the algorithm chooses $\gamma_i < 1$. Combining these observations, we obtain $\phi_i^{k+1, \lambda} \le 0$.

To show complementary slackness, observe that whenever $i$ is not in an ironing interval, as $\gamma_i$ is chosen maximal such that $\overline{ R}^{\kappa(i)}_i =  R_i^{\kappa(i)}$, the Algorithm chooses $\gamma_i= 1$, which implies that for $j > i > r$ and for $j > i+1$ and $i=r$, $\lambda'_{jr} = 0$. By no ironing over maxima, this implies that for $j > i^k > r$, $\lambda_{jr} = 0$. Moreover, as an invariant of the algorithm the flow $\lambda$ is downward. Hence, the only dual variables that are tight are within types that get the same allocation and payment (corresponding to $\lambda_{ij}$ such that $i^k \le i,j \le i^{k+1}$, $k \in [d]$)  or local downward constraints (corresponding to $\lambda_{(i+1)i}$, $i \in [n-1]$). The former incentive constraints clearly bind, the latter bind by weak monotonicity, as for the marginally buying type (which, by compatibility of weak monotonicity with mostly regularity must also be the first type in the MRS order), the price could be raised if she were not indifferent between her allocation and payment and the allocation and payment of the next lower type.
\end{proof}
Having this result, we are ready to finish the proof of \autoref{thm:mrs}. 
\begin{proof}[Proof of \autoref{thm:mrs}]
Implementability follows from weak monotonicity and the definition of the optimal mechanism, \eqref{eq:mechanism2}. Non-negativity and feasibility of flow are properties of the Ironing Algorithm shown in \autoref{lem:feasnonneg}. Virtual welfare maximization and complementary slackness have been shown in \autoref{lem:vwfcsl}. 
\end{proof}

%% file: 5_separate.tex
\section{Upgrade Pricing and Separate Pricing}\label{sec:sep}
In both \autoref{thm:reg} and \autoref{thm:mrs}, we established the optimality of an upgrade pricing mechanism that yields the same allocation as separate (item by item) monopoly pricing, though not necessarily the same transfers. We will show in this section that, under monotonicity with respect to the component-wise partial order, separate pricing and upgrades become equivalent---upgrade pricing is redundant.

We say that the type space $\Theta$ is \emph{monotone} if $\theta_i^k \le \theta_j^k$ for any $i < j \in [n]$ and $k \in [d]$.

We call a mechanism separate pricing if a type separately chooses whether to buy each item $k$ at a price $p_k$. Formally, a mechanism satisfies separate pricing if it can be written as:
\begin{align*}
q_i^k &= \begin{cases} 1 & \theta_i^k \ge p_k\\ 0 & \text{else,} \end{cases} &
t_i &= \sum_{k=1}^d p_k \1_{q_i^k = 1}.
\end{align*}
\begin{theorem}\label{thm:sep}
If the type space $\Theta$ is monotone, then the outcome of any upgrade pricing mechanism can be implemented via separate pricing, and conversely. When the type space is not monotone, neither implication needs to hold.
\end{theorem}
\begin{proof} We first assume types are monotone and show that the allocation and revenue of any upgrade pricing mechanism can be obtained through a separate pricing mechanism, and vice versa.

Let $\theta_1\leq\dots\leq \theta_i\leq\dots\leq\theta_n$, and fix an upgrade pricing mechanism $\mathcal{M}$. This mechanism admits an indirect representation as (a) a collection of bundles ranked by set inclusion $\{b_k\}_{k=0}^K$, with $b_0=\varnothing$ and $K\leq d$, and (b) a vector of prices $t_k$ that are increasing in $k$, with $t_0=0$. Let $\underline{\theta}_k$ and $\bar{\theta}_k$ denote the lowest and highest types who choose bundle $b_k$ under mechanism $\mathcal{M}$. Because types are monotone, buyer self-selection implies $\underline{\theta}_k\geq \bar{\theta}_{k-1}$. 

We now construct a separate pricing mechanism, i.e., a vector of prices $\{p_j\}_{j=1}^d$ that yields the same allocation and payments as our upgrade pricing mechanism. To do so, define the collection of \emph{upgrades} $u_k:=b_k\setminus b_{k-1}$ and the upgrade prices $\tau_k:=t_k-t_{k-1}$. For each upgrade bundle $k$ and every good $j\in u_k$, let the single-item prices $p_j$ satisfy
$$ p_j\in\left[\bar{\theta}_{k-1}^j, \underline{\theta}_{k}^j\right]\quad\text{and}\quad\sum_{j\in u_k}p_j=\tau_k.$$

Under monotonicity, such a vector of prices always exists. By consumer self-selection in  the original mechanism $\mathcal{M}$, we have
\begin{align*}
    \bar{\theta}_{k-1}b_k-t_k&\leq\bar{\theta}_{k-1}b_{k-1}-t_{k-1},\\
    \underline{\theta}_{k}b_{k-1}-t_{k-1}&\leq\underline{\theta}_{k}b_k-t_k.
    \end{align*}
    In turn, this implies $$\bar{\theta}_{k-1}u_k\leq\tau_k\leq\underline{\theta}_{k}u_k.$$

With the prices so constructed, each type purchases the same goods as under $\mathcal{M}$ and pays the same total price. Notice first that each type's choice from the original mechanism $\mathcal{M}$ is still available at the same price, i.e., each bundle $b_k$ can still be purchased for a total price $t_k$. Moreover, by monotonicity, no type $\theta$ who buys bundle $b_k$ under the upgrade pricing mechanism $\mathcal{M}$ derives positive net surplus from any object $j\in u_{k'}$ with $k'>k$ under the separate prices constructed above. And finally, no such type $\theta$ derives positive net surplus by removing any object $j\in u_{k'}$ with $k'\leq k$ from her consumption bundle.

The other direction of this  result is immediate: if types are monotone, the goods purchased by two different types under any separate pricing mechanism are ranked by set inclusion. Thus, replacing the separate pricing mechanism with the resulting upgrade pricing mechanism yields the same outcome.

Finally, we show by means of two counterexamples that, without type monotonicity, separate pricing is not equivalent to upgrade pricing.

In particular, there exist type spaces and vectors of separate prices that do not induce an upgrade pricing allocation. For example, let 
\[
\Theta=\{(1,1),(1,3),(3,3),(4,1)\}
\]
and consider the separate prices $p=(2,2)$: type $\theta_2$ buys  good $2$ only, type $\theta_3$ buys both goods, and type $\theta_4$ buys good $1$ only. 

Likewise, for the same type space, consider the upgrade pricing mechanism where $q=(0,1)$ is sold for $t=2$ and $q=(1,1)$ is sold for $t=4$, i.e., good $j=1$ is only sold as an upgrade, for an additional price $\tau=2$. Under this mechanism, type $\theta_2$ buys good $2$ only, while types $\theta_3$ and type $\theta_4$ buy both goods. However, as we saw above, the vector of separate prices $p=(2,2)$ yields a different allocation (and a lower revenue for the seller). 
\end{proof}
Whenever an upgrade pricing  mechanism implements the allocation of optimal separate pricing, each marginal type $\underline{\theta}_k$ is indifferent by construction between the two consecutive bundles $b_{k-1}$ and $b_k$. Theorem \ref{thm:sep} then implies that the outcome of this mechanism can be implemented by the  separate monopoly prices.
\begin{corollary}\label{cor:mon}
If $\Theta$  is monotone, $q$ is an allocation of an optimal upgrade pricing mechanism, and $q$ is the allocation of separate monopoly pricing, then separate monopoly pricing is optimal.
\end{corollary}
Adding a monotonicity condition to both of our main theorems, \autoref{thm:reg} and \autoref{thm:mrs}, we hence obtain two sets of sufficient conditions under which  separate monopoly pricing is optimal.
\begin{corollary}
If $\Theta$ is monotone and $F$ is regular, separate monopoly pricing is optimal.
\end{corollary}
\begin{corollary}
If $\Theta$ is monotone and has a monotone marginal rates of substitution, and $F$ is mostly regular, then separate monopoly pricing is optimal.
\end{corollary}

%% file: 7_conclusion.tex
\section{Conclusion}\label{sec:conclusion}
It is a common practice for a seller to  offer bundles of products or services that are ordered in a way that more expensive bundles contain all items from less expensive bundles as well as some extra items.
In this paper, we provide sufficient conditions under which such \enquote{upgrade pricing} schemes are exactly optimal for a monopolist seller.

There are several ways in which the current analysis could be extended. First, our conditions could be relaxed to account for richer type spaces and type distributions, such as a continuum of types in the d-dimensional space. One natural extension can be obtained immediately: assume that a type distribution can be split into several type cohorts, in fact quantized type space, such that each type cohort satisfies the conditions of our theorems. Our results imply that the optimal mechanisms in each respective cohort are upgrade pricing. In this respect, \cite{bergemann2021} show that in nonlinear pricing problems, the revenue of the continuous type space is generally well approximated by a finite quantized type space.   

Second, our sufficient conditions for the optimality of upgrade pricing may be complemented by necessary conditions. In doing so, one may want to distinguish between conditions on type distributions and type spaces. For example, one may ask which type spaces guarantee that upgrade pricing is optimal irrespective of the type distribution.

Finally, throughout the paper we highlight the interplay between optimality of different pricing schemes: bundling, upgrade pricing, and separate sales. It would be instructive to  provide a more complete characterization of the cases in which one of these schemes strictly outperforms another.